\documentclass[sn-mathphys-num]{sn-jnl}

\usepackage{graphicx}%
\usepackage{multirow}%
\usepackage{amsmath,amssymb,amsfonts}%
\usepackage{amsthm}%
\usepackage{mathrsfs}%
\usepackage[title]{appendix}%
\usepackage{xcolor}%
\usepackage{textcomp}%
\usepackage{manyfoot}%
\usepackage{booktabs}%
\usepackage{algorithm}%
\usepackage{algorithmicx}%
\usepackage{algpseudocode}%
\usepackage{listings}%
\usepackage{physics, braket, subfloat,graphicx,subfigure,epstopdf,lastpage}

\theoremstyle{thmstyleone}%
\theoremstyle{thmstyletwo}%

\theoremstyle{thmstylethree}%

\begin{document}
	
	\title[Article Title]{Optimizing realistic continuous-variable quantum teleportation with non-Gaussian resources}
	
	
	\author[1]{\fnm{Ankita} }
	\email{ankitapanghal1998@gmail.com}
	
	\author*[1]{\fnm{Arpita} \sur{Chatterjee}}
	\email{arpita.sps@jcboseust.ac.in}

	\affil[1]{\orgdiv{Department of Mathematics}, \orgname{J.C. Bose University of Science and Technology, YMCA}, \orgaddress{ \city{Faridabad}, \postcode{121006}, \state{Haryana}, \country{India}}}
	
	
	\abstract{In this work, we investigate the performance of non-Gaussian entangled resources in continuous-variable quantum teleportation within a realistic setting. We describe the characteristic functions of three distinct entangled resources, a two-mode squeezed vacuum state, a two-mode photon-subtracted squeezed state, and a two-mode photon-added squeezed state. We extend the theoretical analysis by Yang et al. to include the realistic experimental conditions such as photon losses, imperfect measurements which typically affect continuous-variable quantum teleportation. Our results demonstrate that even in non-ideal situations, the photon-subtracted squeezed state outperforms the other two resources in the low squeezing regime, keeping fidelity above the classical threshold that suggests the robustness of photon-subtracted squeezed state in practical teleportation applications. We further analyze the EPR correlations of these entangled resources, revealing that the photon-subtracted squeezed state exhibits stronger EPR correlations than the original two-mode squeezed vacuum state and the two-mode photon-added squeezed state. This study merges theoretical models with realistic imperfections and utilizes non-Gaussian entanglement into high-fidelity quantum teleportation.}
	
	
		
		
		
		
		\maketitle

		\section{Introduction}
		\label{sec1}
		
		Quantum teleportation is a major concept in quantum communication that enables the transfer of quantum states between distant locations without physically transmitting the particle itself. The idea of quantum teleportation is originally demonstrated in discrete-variable systems \cite{P1} and then developed and recognized in continuous-variable (CV) systems \cite{P2} due to its ability to handle quantum states with infinite degrees of freedom, making it highly practical for applications such as quantum networking, quantum cryptography \cite{A1} etc. However, a key challenge in CV quantum teleportation lies in enhancing the fidelity and efficiency of the protocol, especially under realistic conditions where noise and other imperfections are usual. These difficulties become even more notable while dealing with the constraints imposed by decoherence, environmental disturbances and measurement errors.
		
		Traditionally, Gaussian states \cite{A2} have been used as the standard resources for implementing quantum teleportation protocols. They offer simplicity and ease of manipulation which have made them a natural choice in early experiments and theoretical models. However, many research works have highlighted the limitations of Gaussian states, especially in noisy environments. These states are subject to no-go theorems  that prevent optimal error correction \cite{P3} and are less effective to achieve high fidelity in the presence of noise. They also meet significant hurdles in distilling entanglement from mixed states \cite{P4,P5,P6,P7,P8}. These drawbacks and pursuit of high-performance quantum communication protocols have led to an  interest for exploring non-Gaussian states as alternative resources. Several schemes for the generation of non-Gaussian states have been proposed \cite{P8a,P8b,P8c,P8d,P8e,P8f,P8g}, which with their unique non-classical properties, offer substantial advantages over their Gaussian counterparts particularly overcoming the limitations mentioned above.
		
		As non-Gaussian states possess enhanced entanglement, exhibiting negative regions in phase-space distributions and are more resilient to noise, they are in demand for various quantum information applications and quantum estimation tasks. These features have led to recognize the non-Gaussian resources as powerful tools in advancing quantum communication \cite{P9,P10}, quantum computation \cite{P11,P12,P13}, and quantum metrology \cite{P14}. In spite of large number of theoretical works suggesting that the non-Gaussian resources can significantly improve teleportation fidelity \cite{P0,P15,P16,P17,P18,P19}, most of the studies have focused on ideal, noise-free scenarios. However, in practice, teleportation systems must contend with numerous imperfections including measurement errors, losses during transmission, and the effects of environmental noise. These factors introduce significant challenges in teleportation and degrade the quality of the transmitted quantum information. While in ideal case, numerous theoretical studies predict near-perfect teleportation, the real world quantum teleportation works in a much noisy environment. This divergence between theory and practice has motivated us to further investigate how non-Gaussian resources can perform under realistic situation.
		
		Our work provides a comprehensive analysis of the continuous-variable quantum teleportation in presence of noise. We aim to address the robustness of non-Gaussian channels under various decoherence mechanisms such as photon losses, imperfect Bell measurements etc. To achieve this, we employ a systematic framework based on the characteristic function representation, which allows us to extend the ideal teleportation formalism \cite{P0} to account these imperfections. By incorporating multiple sources of noise into our model, we can precisely simulate real-world conditions and evaluate how well non-Gaussian resources can perform in practical quantum systems. 
		
		In particular, our study focuses on optimizing the performance of CV quantum teleportation in noisy channels. These optimizations are crucial for enhancing the fidelity and success rate of teleportation protocols, especially when non-Gaussian resources are used. Through this approach, we aim to bridge the gap between theoretical prediction and practical implementation, ultimately contributing to the development of more reliable and efficient quantum communication networks.
		The paper is organized as follows: in Section \ref{Sec2}, we explain a basic strategy of teleportation and describe the entangled resource states which we are going to use. We calculate the EPR correlation for these entangled resources in Section \ref{Sec3}. In the next section, we extend the idea of characteristic function in realistic scenario and then calculate the fidelity for teleporting an input coherent state. Finally we draw our conclusion based on the results obtained. 
		\section{Non-Gaussian resources}
		\label{Sec2}
		In this section, we present a realistic implementation of the continuous-variable Braunstein-Kimble (BK) teleportation protocol by using the characteristic function formulation. This approach is particularly useful for modelling realistic scenarios \cite{P19,P20,P21,P22,P23,P24}, where non-ideal factors come into play. In the protocol, Alice and Bob share a two-mode entangled state. Alice, located faraway from Bob, wishes to teleport a quantum state $\ket{\phi}$ to him. For this, Alice first performs a homodyne measurement on her part of the entangled channel along with the state $\ket{ \phi}$ she intends to transmit. This measurement is performed by using a 50:50 beam splitter, and the outcome of the measurement is communicated to Bob via a classical channel. Based on the information Alice sent, Bob applies a unitary displacement operator to his part of the entangled state that effectively completes the teleportation.
		
		In a realistic set-up, there are several imperfections that need to be considered. During the transmission, Bob's qubit may experience decoherence due to environmental interactions over long distance. Alice's homodyne measurement is also subject to practical limitations such as photon detection inefficiency and background noise which can distort the measurement outcomes. Thus the output teleported state is affected by the inefficiency of the homodyne detector and the decoherence rate of the noisy channel. To model this, we begin by using the two-mode squeezed vacuum state as the entangled resource. This state is a fundamental element in many quantum information processing tasks such as teleportation \cite{P29}, dense coding \cite{P30}, and entanglement swapping \cite{P31,P32,P33} due to its strong entanglement properties. The two-mode squeezed vacuum state can be obtained by applying the two-mode squeezing operator on vacuum state as follows:
		\begin{align} 
			\ket{\psi_{\text{squ}}}  &= e^{r(a^\dagger b^\dagger-ab)}\ket{0,0} = \sqrt{1-\lambda^2}\sum_{n=0}^\infty \lambda^n\ket{n,n} 
			\label{1}
		\end{align}
		where $\lambda = \tanh{r}$ with the squeezing parameter $r$ that ranges between 0 and 1, $a$ $(a^\dagger)$ and $b$ $(b^\dagger)$ are the photon annihilation (creation) operators for mode 1 and 2, respectively. Although the two-mode squeezed vacuum state provides a powerful resource for quantum teleportation, we cannot overlook the potential advantages of using the non-Gaussian states that can be generated by adding (subtracting) photons to (from) a two-mode squeezed vacuum state.
		
		\subsection{Photon-subtracted squeezed vacuum state}
		By subtracting a single photon from each mode of the two-mode squeezed vacuum state, the non-Gaussian photon-subtracted squeezed vacuum state is obtained as \cite{P0}
		\begin{align} 
			\ket{\psi}_s &= N_sab \ket{\psi_{\text{squ}}} = \sqrt{\frac{(1-\lambda^2)^3}{1+\lambda^2}}\sum_{n=0}^\infty \lambda^{n}(n+1)\ket{n,n} 
			\label{2}
		\end{align}
		where $N_s$ is the normalization constant. In experiment, this photon-subtracted state can be generated from two-mode squeezed state by using a beam splitter of low reflectivity \cite{P34}.
		\subsection{Photon-added squeezed vacuum state}
		Similarly by adding a single photon to each mode, we can get the photon-added squeezed vacuum state as \cite{P0}
		\begin{align} 
			\ket{\psi}_a &= N_a a^\dagger b^\dagger \ket{\psi_{\text{squ}}} = \sqrt{\frac{(1-\lambda^2)^3}{1+\lambda^2}}\sum_{n=0}^\infty \lambda^{n}(n+1)\ket{n+1,n+1}
			\label{3}
		\end{align}	
		where $N_a$ is the normalization constant. This state can be manufactured experimentally by using a parametric down converter with low gain \cite{P35}. 
		
		These non-Gaussian states exhibit non-classical features which may enhance their entanglement and correlation characteristics, offering better possibility for quantum information processing tasks. Thus we explore the potential of considered non-Gaussian states to produce and manipulate quantum correlation by using the Einstein-Podolsky-Rosen (EPR) relation.	
		\section{EPR correlation}
		\label{Sec3}
		Quantum correlation is a key ingredient for realizing continuous-variable quantum teleportation \cite{P29}. In Braunstein and Kimble protocol, the quantum channels rely on the EPR correlations and the fidelity for teleporting an arbitrary input state is also governed by them. Here the non-Gaussian resources can be characterized by the EPR correlations between phase-space quadrature components of the two modes. The phase-space quadrature operators of each mode are defined as $ X_j = \frac{1}{\sqrt{2}} (a_j+a_j^{\dagger} )$ and  $ P_j = \frac{1}{ \iota \sqrt{2}} (a_j-a_j^{\dagger} )$ where $j=1, 2$. The EPR correlations of non-Gaussian states studied in this work provide essential insights into their quantum properties and affirm their suitability for continuous-variable teleportation protocols. 
		
		 In the vacuum state, both the variances $\triangle(x_1 - x_2)^2 \ $ and $ \triangle(p_1 + p_2)^2 $ are equal to 1. For any classical two-mode state, both of these exceed 1, reflecting the absence of strong correlations between the modes. However, in the Einstein-Podolsky-Rosen (EPR) state \cite{35}, $\triangle(x_1 - x_2)^2  = \triangle(p_1 + p_2)^2  = 0$. This condition remarkably implies that the quadratures $x_1$ and $p_1$ of the first mode can be perfectly predicted by measurements of $x_2$ and $p_2$ of the second mode or vice versa, implying the existence of ideal EPR correlation between the two modes. This ideal correlation is an essential feature of quantum mechanics and challenges the notion of local realism. Interestingly, a two-mode squeezed state can exhibit stronger EPR correlations beyond the limit of the vacuum state. Whenever the squeezing parameter is not equal to zero, both the variances $\triangle (x_1 - x_2)^2 $  and  $\triangle(p_1 + p_2)^2$ are less than 1, indicating enhanced entanglement and more precise correlations between the two modes. Such states are of significant interest for applications in quantum communication, quantum teleportation, and quantum information processing where stronger correlations can improve the fidelity and thus the efficiency of quantum protocols.
		
		For a two-mode squeezed vacuum state, the EPR correlation 
		\begin{align}		
		\triangle (x_1 - x_2)^2 = 1- \frac{2\lambda}{1+\lambda}  
			\label{eq4}
		\end{align}
directly depends on the squeezing parameter $\lambda$ $(=\tanh {r})$ and is below the vacuum state limit for $\lambda>0$. For the photon-subtracted ($\ket{\psi}_{\text{s}}$) and photon-added ($\ket{\psi}_{\text{a}}$) squeezed vacuum states, the respective variances are
		\begin{align}
		{ \triangle(x_1 - x_2)_s^2 = 1-\frac{4\lambda(\lambda^2-\lambda+1)}{(1+\lambda^2)(1+\lambda)} } ,
			\label{eq5}
		\end{align}
				and
				\begin{align}
		 { \triangle(x_1 - x_2)_a ^2 = 1- \frac{2(\lambda^3-\lambda^2+3\lambda-1)}{(1+\lambda^2)(1+\lambda)} } ,
			\label{eq6}
		\end{align}
		For all the states \eqref{1}-\eqref{3}, $\triangle(x_1 - x_2) ^2=\triangle(p_1 + p_2)^2$. These expressions highlight how the squeezing parameter influences the strength of the EPR correlations. 		 
			\begin{figure*}[h]
			\centering
		\includegraphics[width=0.45\columnwidth]{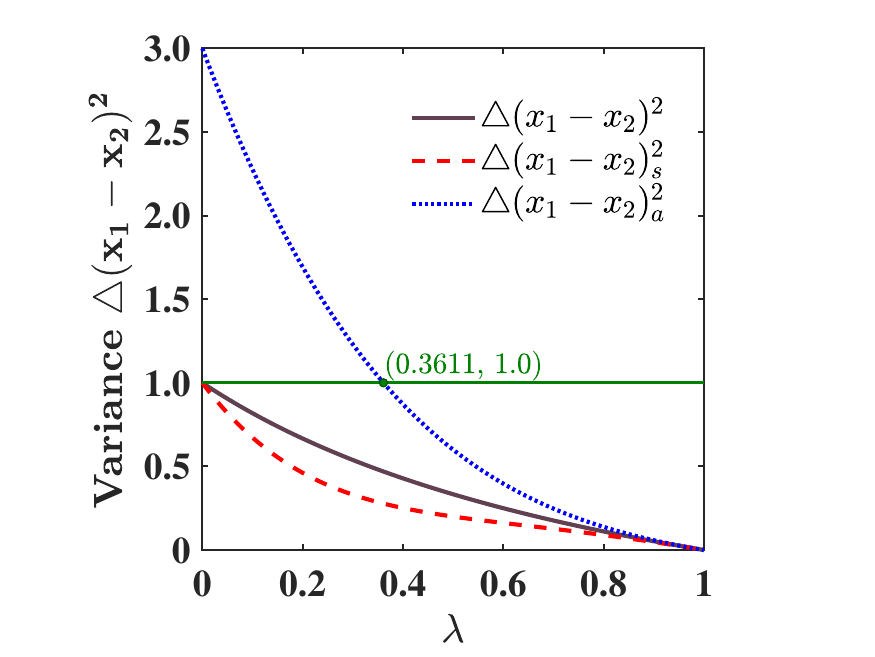}
			\caption{(Color online) Variance $\triangle (x_1 - x_2)^2 $ as a function of $\lambda$ for the two-mode squeezed state (black solid line), photon-subtracted squeezed vacuum state (red dashed line) and photon-added squeezed vacuum state (blue dotted line).}
			\label{fig1}
		\end{figure*} 
		
The plot of EPR correlations provides a clear understanding of how these correlations vary with the squeezing parameter \(\lambda\). In Fig.~\ref{fig1}, we can see that whenever squeezing parameter is greater than $0$, the EPR correlations of the two-mode squeezed state and the photon-subtracted squeezed state go beyond the limit of the vacuum state which is 1. But the EPR correlation of the photon-added squeezed state is initially higher than 1, shows steeper changes and indicates stronger quantum correlations when $\lambda\geq 0.36$ (approximately). As $\lambda$ increases, all the correlations approach the ideal zero-valued EPR correlation  between the two modes. It is evident from the figure that higher squeezing value results stronger EPR correlations for all the three states. Also the EPR correlation can be enhanced by using the photon subtraction process. Moreover, in the BK protocol of CV quantum teleportation, the quantum channel is based on the EPR correlations and the fidelity of teleported states is determined
by the EPR correlations. Thus, one may expect that the quality of quantum teleportation can be
improved by use of either the photon-subtracted state or the
squeezed vacuum state as entangled resource. In fact, Dell'Anno et al. \cite{P18} showed that the CV teleportation fidelity is enhanced by using the photon-subtracted state as quantum channel. These observations emphasize the trade-off between entanglement strength and noise sensitivity, guiding the choice of resource states for realistic quantum teleportation set-ups.
		
		After understanding the EPR correlations, we examine how these correlations can enhance the quality of teleportation protocol. Fidelity is a key measure that reflects how well the teleported state matches with the original one. To calculate the fidelity, the characteristic function formalism is used which provides a straightforward and reliable framework. In the following section, we calculate the characteristic functions for three different states to analyze the effectiveness of the teleportation process.  
		\section{Characteristic function}
		\label{Sec4}
		In quantum mechanics, the characteristic function provides a fundamental way to describe quantum states in terms of quasi-probability distribution in phase-space. It enables the calculation of various important quantities such as Wigner function, expectation values and fidelity between quantum states. By utilizing the phase-space formalism, the characteristic function becomes particularly important while dealing with non-Gaussian states and complex quantum correlations. 
		
		Let us assume that the density matrices for any single-mode pure input state and two-mode pure entangled resource are $\rho_{\text{in}}=\ket{\phi}_{\text{in} \ \text{in}} \! \bra{\phi}
		$ and $\rho_{\text{res}}=\ket{\psi}_{12 \ 12} \! {\bra{\psi}}$, respectively. Initially, the input state is not entangled with any qubit of resource. Therefore the initial three mode field is given as $\rho_0 = \rho_{\text{in}} \otimes \rho_{\text{res}}$ and their characteristic function is \cite{P36} 
		\begin{align*}
			\chi_0(\gamma_{\text{in}};\gamma_1;\gamma_2) &= Tr[\rho_0 D_{\text{in}}(\gamma_\text{in}) D_{1}(\gamma_{1})D_{2}(\gamma_{2})] \\
			&= \chi_{\text{in}}(\gamma_{\text{in}})  \chi_{\text{res}} (\gamma_{1};\gamma_{2}) 
		\end{align*}
		where $Tr$ denotes the trace operation, $ D_{j}(\gamma_{j}) $ is the displacement operator for the mode $j$ $(j=\text{in}, 1, 2)$, $\chi_{\text{in}}$ and $\chi_{\text{res}}$ are the characteristic functions for input and resource states respectively. As the phase-space quadrature operators are defined as $ X_j = \frac{1}{\sqrt{2}} (a_j+a_j^{\dagger} )$ and  $ P_j = \frac{1}{ \iota \sqrt{2}} (a_j-a_j^{\dagger} )$ where $j=\text{in}, 1, 2$ and the corresponding real phase-space variables are $ x_j = \frac{1}{\sqrt{2}} (\gamma_j+\gamma_j^{\dagger} )$ and  $ p_j = \frac{1}{ \iota \sqrt{2}} (\gamma_j-\gamma_j^{\dagger} )$, the characteristic function can be written in terms of the quadrature variables as $ \chi_0(\gamma_{\text{in}};\gamma_1;\gamma_2) =\chi_0(x_{\text{in}},p_{\text{in}};x_{1},p_{1};x_{2},p_{2}) $.
		
		For any single-mode input state, the combined effect of propagation through a damping channel, application of a unitary displacement, and a non-ideal homodyne measurement determines the characteristic function of the final output state $\chi_{\text{out}}(x_2,p_2)$ as follows \cite{P36}
		\begin{align}\nonumber
			\chi_{\text{out}}(x_2,p_2) &= \chi_{\text{in}}(gT\gamma) 	\chi_{\text{res}} (gT\gamma^*;e^{-\frac{\tau}{2}}\gamma) 
			\exp(-\frac{1}{2} \Gamma_{\tau,R}|\gamma|^2)\\\nonumber
			&= \chi_{\text{in}}(gTx_2,gTp_2) 
			\chi_{\text{res}} (gTx_2,-gTp_2;e^{-\frac{\tau}{2}}x_2,e^{-\frac{\tau}{2}}p_2) \cross \nonumber  \\ & 
			\exp(-\frac{1}{2} \Gamma_{\tau,R}(x_2^2+p_2^2)) 
			\label{4}
		\end{align}
		Here $g$ denotes the gain factor, $\tau = \gamma t$ where $\gamma$ is the damping rate and $\Gamma_{\tau,R}$ represents the thermal phase-space covarience given by
		\begin{align}
			\Gamma_{\tau,R} = (1-e^{-\tau})\left(\frac{1}{2}+n_{\text{th}}\right) +g^2R^2,
		\end{align}
		$T$ and $R$ are the transmissivity and reflectivity of the beam splitter with $ T^2+R^2=1 $. The effect of imperfect Bell measurement is evidenced by the scaling factor $T$ in the arguments of the input and mode 1 of the resource characteristic functions, $\chi_{\text{in}}$ and $\chi_{\text{res}}$ respectively.  The decoherence caused by noisy propagation affects only mode 2 of the resource state and is represented by the exponentially decreasing factor $e^{-\frac{\tau}{2}} $ in the argument of $\chi_{\text{res}}$. The output characteristic function for the ideal teleportation can be obtained from \eqref{4} when $R=0$ $(T = 1)$, $\gamma=0$ $(\tau=0)$ and $g =1$ as
		\begin{align}
			\chi_{\text{out}}(x_2,p_2) &= \chi_{\text{in}}(x_2,p_2) 
			\chi_{\text{res}} (x_2,-p_2;x_2,p_2) 
		\end{align}
		The characteristic function of the two-mode squeezed vacuum state \cite{P0} is obtained as
		\begin{align}
			\chi(\alpha,\beta) &= \exp\left(-\frac{1+\lambda^2}{2(1-\lambda^2)}(|\alpha|^2 + |\beta|^2)   + \frac{\lambda}{(1-\lambda^2)}(\alpha\beta + \alpha^* \beta^*)\right) 
			\label{7}
		\end{align} 
	and the characteristic functions for the photon-subtracted and photon-added squeezed states are \cite{P0} 
		\begin{align}
			\chi_s(\alpha,\beta) &= N_s^2 \exp(-(|\alpha|^2 + |\beta|^2)/2) \wedge_s(\alpha)\wedge_s\beta)\left[\chi(\alpha,\beta)\exp((|\alpha|^2 + |\beta|^2)/2) \right]  \label{8}
		\end{align}
		and
		\begin{align}
			\chi_a(\alpha,\beta) &= N_a^2 \exp(-(|\alpha|^2 + |\beta|^2)/2) \wedge_a(\alpha)\wedge_a(\beta)\left[\chi(\alpha,\beta)\exp((|\alpha|^2 + |\beta|^2)/2) \right] \label{9}
		\end{align}
		where the operator $\wedge_s(\alpha) $ and $\wedge_a(\alpha)$ are given as 
		\begin{align*}
			\wedge_s(\alpha) &=	\frac{\partial}{\partial \alpha} \frac{\partial}{\partial \alpha^*} \\
			\wedge_a(\alpha) &=	-\frac{\partial}{\partial \alpha} \frac{\partial}{\partial \alpha^*}+\alpha\frac{\partial}{\partial \alpha}+\alpha^*\frac{\partial}{\partial \alpha^*}-\alpha\alpha^*+1
		\end{align*}
		and for $ \wedge(\beta)$, $ \alpha $ is replaced by $ \beta $. In the next section, we compute and analyze the fidelities for teleporting an input coherent state via three different entangled channels $\ket{ \psi_{\text{squ}}}$, $\ket{ \psi}_s$ and $\ket{ \psi}_a$.
		\section{Realistic teleportation with non-Gaussian resources}
		\label{Sec5}
		Quantum teleportation is a fundamental protocol in quantum information theory that allows the transfer of an unknown quantum state between two distant parties without physically transmitting the state itself.
		The success of a teleportation protocol is measured by the fidelity which quantifies how closely the state reconstructed at the receiving end matches with the original state under consideration. The states are typically represented by phase-space variables, and the fidelity can be computed by examining the overlap between the characteristic functions of the input and output states. Fidelity, ranging between $0$ to $1$, measures the overlap between the original state and the reconstructed state, where a fidelity value 1 corresponds to perfect teleportation and a fidelity value 0 represents complete failure. Under the BK protocol, perfect teleportation fidelity
($F = 1$) is obtained with an infinitely nonclassical channel
such as the ideal EPR entangled state \cite{35}. In continuous-variable teleportation, a fidelity value $2/3$ is considered as the no-cloning limit \cite{himadri}. To ensure that the teleported state is the best
copy of the state remaining after the protocol and the nonclassical features of the input state have been teleported,
the average fidelity must be greater than the no-cloning limit.
Hence $F_{\text{clone}}= 2/3$ is an important benchmark for the success of the protocol.

		We now derive the general expression of fidelity for teleporting a single-mode coherent state by way of three different resources $\ket{\psi}$, $\ket{\psi}_s$ and $\ket{\psi}_a$. In the characteristic function description, the fidelity for continuous-variable quantum teleportation is given by \cite{P0} \begin{align}
			F&=\frac{1}{\pi}\int d^2\gamma\, \chi_{\text{in}}(\gamma)\chi_{\text{out}}(-\gamma)
			\label{10}
		\end{align} 
		where $\chi_{\text{in}}(\gamma)$ and  $\chi_{\text{out}}(\gamma)$ are the characteristic functions for the input and the teleported states respectively. Using \eqref{4}, we get 
		\begin{align}
			F&=\frac{1}{\pi}\int d^2\gamma \chi_{\text{in}}(\gamma)\chi_{\text{in}}(-gT\gamma) 	\chi_{\text{res}} (-gT\gamma^*;-e^{-\frac{\tau}{2}}\gamma) 
			\exp(-\frac{1}{2} \Gamma_{\tau,R}|\gamma|^2)
			\label{11}
		\end{align} 
		The characteristic function $\chi_{\text{in}}$ for the single-mode input coherent state $\ket{\alpha}$ value of  that is the  can be calculated easily using $\chi{(\gamma)} = \text{Tr}[\rho D(\gamma)]$ where $\rho$ is its density matrix as
		\begin{align}
			\chi_{\text{in}}(\gamma) &=\exp(-\frac{|\gamma|^2}{2})\exp(\alpha^\star\gamma-\alpha\gamma^\star)
			\label{12}
		\end{align}
		By substituting \eqref{12} and \eqref{7}-\eqref{9} into \eqref{11}, the fidelities for squeezed vacuum, photon-subtracted squeezed vacuum and photon-added squeezed vacuum resources can be computed as
		\begin{align}
			F	&= \frac{1}{2S}\exp(-\frac{|\alpha|^2P^2}{S}) \label{13} \\
			F_s	&=\frac{N_s^2}{2S}   \exp(-\frac{|\alpha|^2P^2}{S})\Bigg[  (A^2 + B^2)     	   
			+ ( A G^2  +2 BGH   + AH^2) \cross \nonumber \\
			&\quad \frac{1}{S} \left(1-\frac{|\alpha|^2P^2}{S} \right) 
			+G^2 H^2    \frac{1}{S^2} \left(2-\frac{4|\alpha|^2P^2}{S}  +\frac{P^4|\alpha|^4}{S^2}\right)   \Bigg] \label{14} \end{align}
		and
		\begin{align}	
			F_a 	&=\frac{N_a^2}{2S}   \exp(-\frac{|\alpha|^2P^2}{S})\Bigg[  L    	   
			+ M \frac{1}{S} \left(1-\frac{|\alpha|^2P^2}{S}\right)     
			+N  \frac{1}{S^2}\cross  \nonumber \\ &\quad    \left(2-\frac{4|\alpha|^2P^2}{S} +\frac{P^4|\alpha|^4}{S^2}\right)   \Bigg]  \label{15}
		\end{align}
		with
		\begin{align*}	
			S &=\frac{1}{4} + \frac{g^2T^2}{4} + \frac{(1+\lambda^2)(g^2T^2+e^{-\tau})}{4(1-\lambda^2)} - \frac{\lambda gTe^{-\tau/2}}{(1-\lambda^2)}  + \frac{\Gamma_{\tau,R}}{4}  \\ 
			P &= \frac{1-gT}{\sqrt2},  \,
			G= A\frac{gT}{\sqrt{2}} + B\frac{e^{-\tau/2}}{\sqrt{2}},\,  H = A\frac{e^{-\tau/2}}{\sqrt{2}} + B\frac{gT}{\sqrt{2}}  \\
			A &=-\frac{\lambda^2}{(1-\lambda^2)}, \, B=\frac{\lambda}{(1-\lambda^2)},\, L=	 
			(A-1)^2 + B^2 \\
			M &= (g^2T^2 +e^{-\tau}) \left\{\frac{(A-1)^3}{2}+\frac{3B^2(A-1)}{2}\right\} +gTe^{-\tau/2}\left\{3B(A-1)^2+B^3\right\}
			&\quad \\
			N   &=   (gTe^{-3\tau/2}  + g^3T^3e^{-\tau/2} ) \left\{\frac{B(A-1)}{2}[(A-1)^2+B^2]\right\} + g^2T^2e^{-\tau}\cross \\
			&\quad    \left\{\frac{(A-1)^4}{4}+B^2(A-1)^2+\frac{B^4}{4}   \right\}+( g^4T^4+e^{-2\tau} )\frac{B^2(A-1)^2}{4} 
		\end{align*}
		Here the fidelity is calculated in a realistic scenario accounting the real-world constraints and limitations. In order to understand the decoherence mechanism, we consider the following two cases: (i) decoherence due to imperfect Bell measurement for which we have fixed $\tau = 0$ and considered different values of $R$; (ii) decoherence associated with propagation through noisy channel for which we have fixed $R^2$ (=0.5) and varied the factor $\tau$. 
		\begin{figure}[htbp]
			\centering
			\includegraphics[width=\columnwidth]{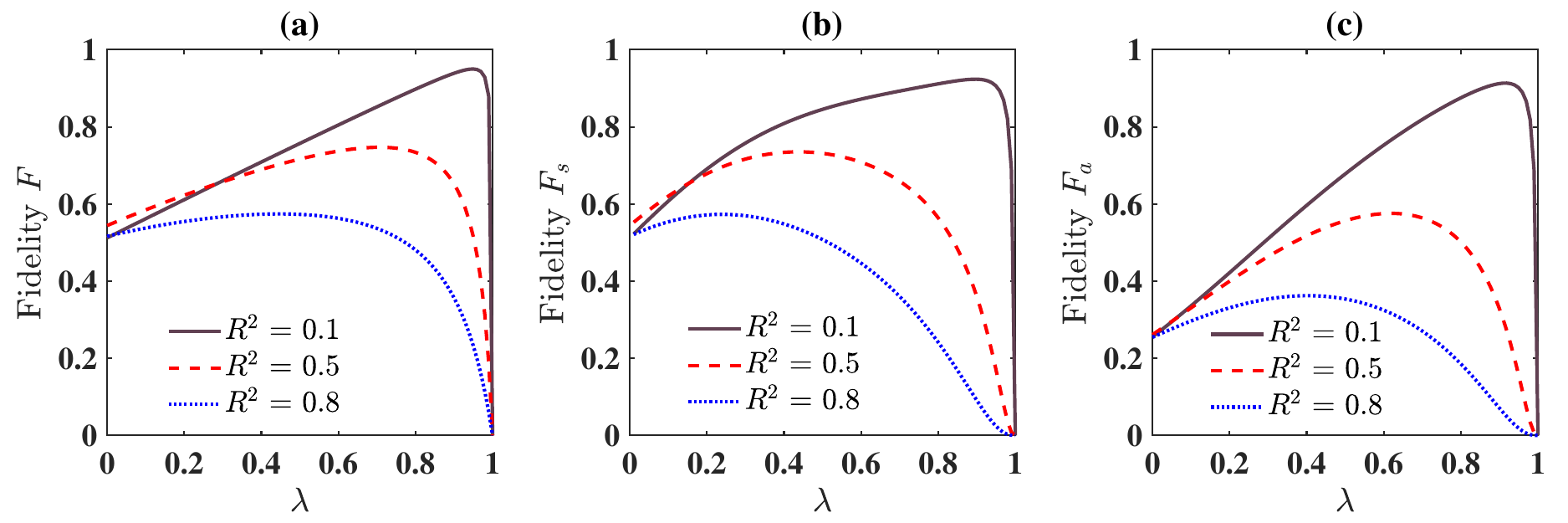}
			\caption{(Color online) Fidelity for teleporting a coherent state $\ket{\alpha}$ as a function of squeezing parameter $\lambda=\tanh{r}$ with $g=1$, $\alpha=1$, $\tau=0$, $n_{\text{th}}=0$ and for (a) two-mode squeezed vacuum state, (b) photon-subtracted squeezed state and (c) photon-added squeezed state. The reflectivity of the homodyne measurement is differed as $R^2=0.1$ (black solid line), $R^2=0.5$ (red dashed line) and $R^2=0.8$ (blue dotted line) in each panel.}
			\label{fig2}
		\end{figure}  
		\begin{figure*}[htbp]
			\centering
			\includegraphics[width=\columnwidth]{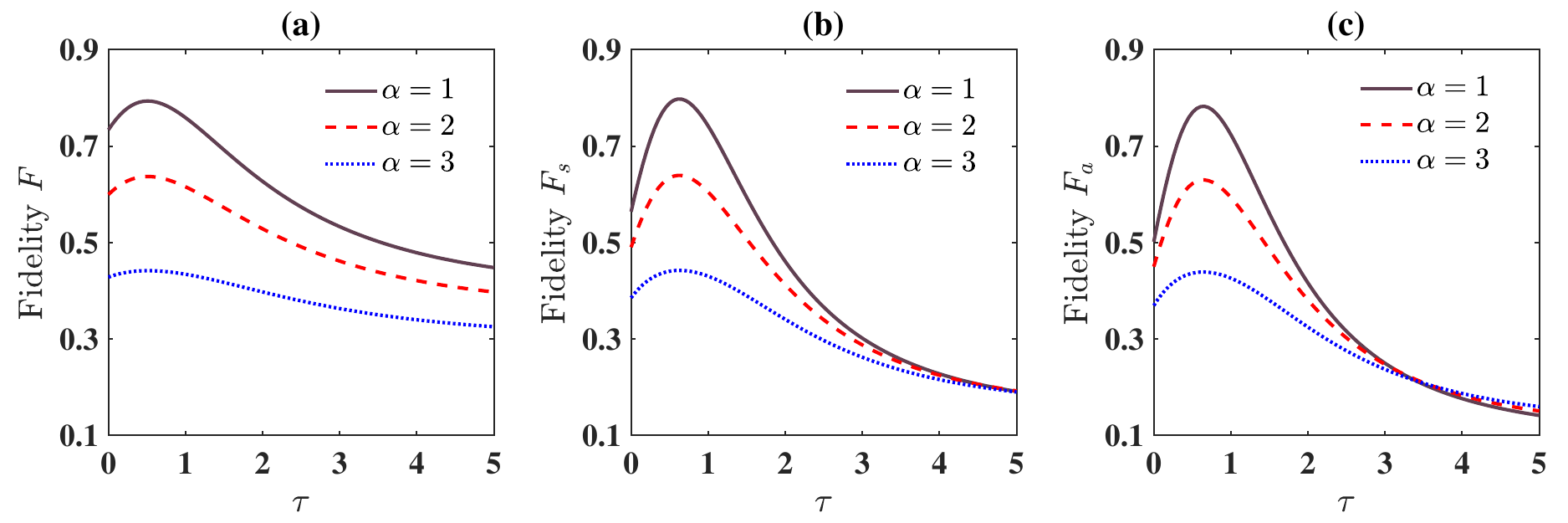}
			\caption{(Color online) Fidelity for input coherent state $\ket{\alpha}$ as a function of reduced time $\tau$ with $g=1$, $R^2=0.5$, $n_{\text{th}}=0$ and for (a) two-mode squeezed vacuum state (b) photon-subtracted squeezed state and (c) photon-added squeezed state. Here $\alpha =1$ (solid black line), $\alpha =2$ (red dashed line), $\alpha =3$ (blue dotted line) in all the three cases.}
			\label{fig3}
		\end{figure*}
		
		Fig.~\ref{fig2} shows the variation of fidelity with respect to the squeezing parameter $\lambda$. We have fixed realistic parameter values such as $g=1$, $\alpha=1$, $\tau=0$, $n_{\text{th}}=0 $ and considered different values of the reflectivity parameter, $R^2 = 0.1,\,0.5$ and $0.8$, to understand the effect of imperfect Bell measurement on fidelity. We can clearly see that the non-ideal protocol works best for photon-subtracted squeezed state resource which is also true in ideal case. In higher squeezing regime, the maximum average fidelity surpasses the threshold of cloning fidelity. It is worth noting that for $R^2 = 0.8$, the maximum fidelity is achieved at very low values of the squeezing parameter ($\lambda\approx 0$).
				
		Fig.~\ref{fig3} describes the changes in fidelity for sending a single-mode coherent state $\ket{\alpha}$ across three different entangled channels. In order to understand the effect of noisy environment during propagation, the realistic parameters are fixed at $g=1$, $R^2=0.5$, and $n_{\text{th}}=0$. We have observed that the fidelity decreases over time in all the cases and the decline is more rapid for both the photon-subtracted and photon-added states compared to the two-mode squeezed vacuum state. This is attributed by the higher sensitivity of photon-subtracted and photon-added states to decoherence and noise which accelerates their fidelity loss. \\
		 Next, we have calculated the fidelities for the same resources under ideal conditions that provides a comparison between ideal and realistic outcomes. The output fidelity for the ideal teleportation can be obtained from \eqref{13}-\eqref{15} by putting $R=0$ $(T = 1)$, $\gamma=0$ $ (\tau=0)$ and $g =1$ 
		\begin{align}
			F^{\text{ideal}}	&= \frac{1+\lambda}{2}  \\
			F^{\text{ideal}}_s		&= \frac{(1+\lambda)^3(\lambda^2-2\lambda+2)}{4(1+\lambda^2)} \\
			F^{\text{ideal}}_a &=  \frac{(1+\lambda)^3}{4(1+\lambda^2)}
		\end{align} 
		These results can also be verified from Yang's work \cite{P0}.	
		\begin{figure*}[htbp]
			\centering
			\includegraphics[width=0.45\columnwidth]{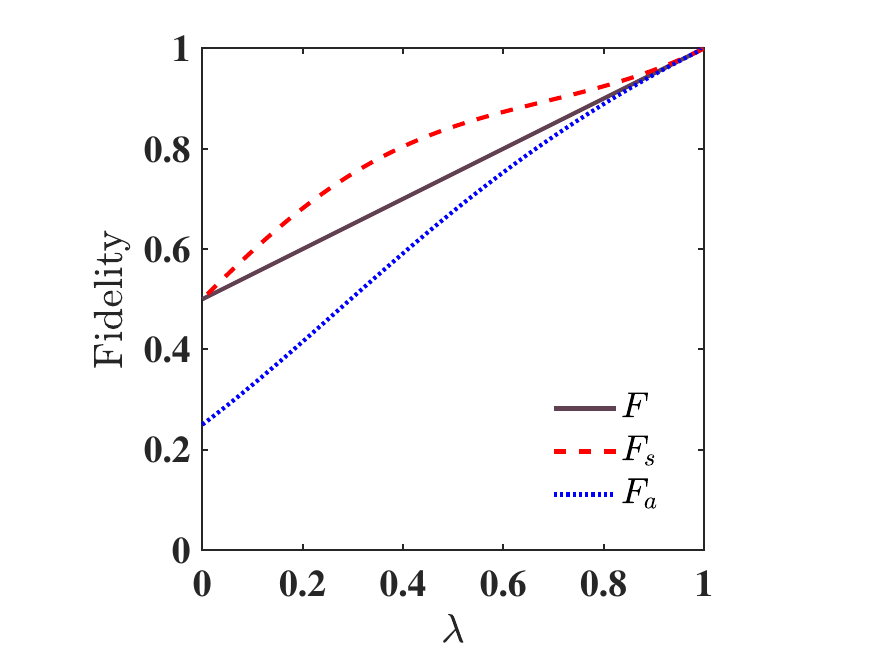}
			\caption{(Color online) Teleportation fidelity for input coherent state $\ket{\alpha}$ as a function of $\lambda$ with $g=1$, $\alpha=1$, $\tau=0$, $R^2= 0$ and $n_{\text{th}}=0$ for the following entangled resources: two-mode squeezed state (full line), photon-subtracted squeezed state (dashed line) and photon-added squeezed state (dotted line).}
			\label{fig4}
		\end{figure*} 
		For a comparison of the resources under ideal conditions, Fig.~\ref{fig4} is plotted. We can clearly see that in low squeezing regime, the photon-subtracted squeezed state is the best resource for teleportation. This work suggests that photon-subtracted squeezed state could significantly improve the performance of quantum teleportation, both in practical scenario where imperfections and losses are inevitable as well as in ideal situation.
		\section{Conclusion}
		\label{Sec6}
		In this paper, the entanglement properties of non-Gaussian states are investigated which are obtained by adding (subtracting) photons to (from) a two-mode squeezed vacuum state. The EPR correlations between the phase-space quadrature components of the two modes are calculated. We have found that both the original two-mode squeezed vacuum state and the photon-subtracted squeezed state exhibit stronger EPR correlations. We have studied the Braunstein-Kimble protocol for teleporting a single-mode coherent state by means of non-Gaussian resource channels in realistic scenario. We have also examined how these resources perform in ideal situation. The non-ideal protocol is subject to the decoherence effects such as photon losses in optical fibre and imperfections in Bell measurement. Using the characteristic function formalism, we have discussed the impact of decoherence on the performance of different non-Gaussian states. Our results show that, despite the challenges offered by the decoherence effects, the fidelity associated with different quantum resources remains above the classical benchmark. Notably, the photon-subtracted squeezed vacuum state outperforms other two states both in ideal and non-ideal teleportation protocols. 
		
		\section{Acknowledgements}
		Ankita's work is supported by the University Grants Commission (UGC), Govt. of India (Award no. 231610110670). A. C. acknowledges DST SERB for the support provided through the project number SUR/2022/000899.

	\end{document}